\documentstyle[11pt,paspconf,epsf]{article}
\def\Msun{{\rm\,M_\odot}}
\def\Lsun{{\rm\,L_\odot}}

\begin{document}
\title{Evidence for a Massive Black Hole in the S0 Galaxy NGC~4342}

\author{Nicolas Cretton}
\affil{Sterrewacht Leiden, The Netherlands}
\author{Frank C. van den Bosch}
\affil{University of Washington, Seattle, USA}

\begin{abstract}
We have constructed axisymmetric dynamical models of the edge-on S0
galaxy NGC 4342: simple two-integral Jeans models as well as fully
general, three-integral models using a modified version of
Schwarz\-schild's orbit superposition technique.  The two-integral
models suggest a black hole (BH) of 3 or $6\times 10^8 {\rm M_\odot}$,
depending on the data set. The three-integral models can fit all
ground-based and HST data simultaneously, but only when a central BH
is included. Models without BH are ruled out at better than
99.73\% confidence level. We determine a BH mass of $3.0^{+1.7}_{-1.0} \times
10^8 \Msun$. This corresponds to $2.6\%$ of the bulge mass, making NGC 4342
one of the galaxies with the highest BH mass to bulge mass ratio
currently known.
\end{abstract}

\section{The data}

NGC 4342 is a low luminosity ($M_B = -17.47$) edge-on S0 galaxy 
in the Virgo cluster, displaying both an outer disk ($r > 5''$)
 and a stellar nuclear disk ($r < 1''$). 

 The data consists of long slit spectroscopy obtained from the ground,
 and of multi-color photometry and single aperture spectroscopy
 obtained with HST (see van den Bosch, Jaffe \& van der Marel 1998).
 We modelled this complex mass distribution using the Multi Gaussian
 Expansion technique of Emsellem, Monnet \& Bacon (1994).

\section{Jeans models}

We have first constructed simple dynamical models assuming a
distribution function (DF) of the form ${\rm DF} = f(E,L_z)$.  Under
the assumption of constant mass to light ratio (M/L), we have solved
the Jeans equations and compared the projected velocity dispersions
(after seeing convolution and pixel binning) to the observations.
Although this model shows some discrepancies with the data, it
suggests a BH mass in the range (3 - 6)$\times 10^8 \Msun$ (see
Cretton \& van den Bosch 1998 for details). These models cannot
 fit simultaneously the two data sets, suggesting a more sophisticated 
type of modeling (three-integral DF).

\begin{figure}
\vskip-3cm
\plotone{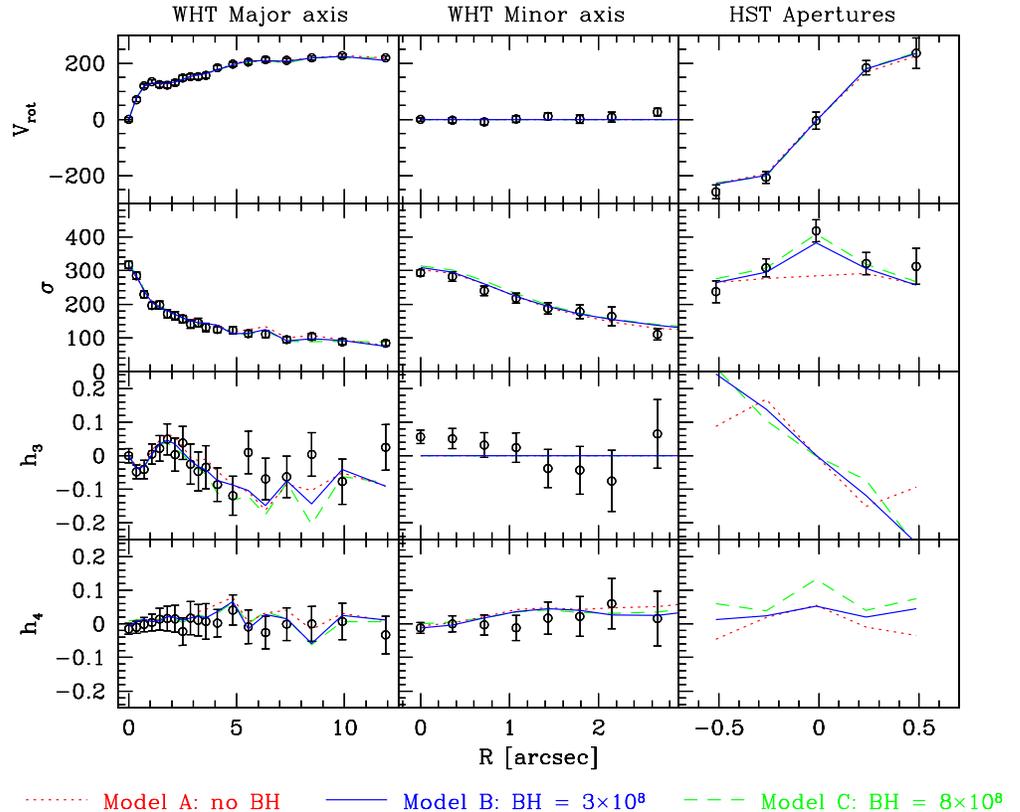}
\caption{Fits to the kinematic data for NGC 4342: mean streaming velocity $V$,
velocity dispersion $\sigma$ and Gauss-Hermite moments $h_3$ and $h_4$
for ground-based long slit data (William Herschel Telescope) and HST
FOS apertures. Model A (dotted line) has no BH, model B (solid
line) a $3\times 10^8 \Msun$ BH and model C (dashed line) a
$6\times 10^8 \Msun$ BH.}
\end{figure}

\section{Three-integral models}

 Our scheme follows Schwarzschild's method with an extension towards
 the use of kinematic constraints: We first deproject the surface
 brightness to obtain a three-dimensional mass density distribution
 (assuming a constant M/L). From the Poisson equation the gravitational
 potential corresponding to the stellar body is derived. At that
 stage, a dark component (BH or dark halo) can be added to the
 potential. In this total potential, a large library of orbits is
 computed and projected onto the observable space (position on the
 sky, line-of-sight velocities).  Again the necessary steps of seeing
 convolution and pixel binning are performed for a fair comparison to
 the data. Finally the orbital weights are computed that best
 reproduce the photometric and kinematic constraints (see Figure 1 for
 the kinematic fits of 3 models).

The inclusion of the complete distribution of line-of-sight velocities
 greatly helps in constraining the final model. The three-integral
 models have more freedom than the simpler Jeans models and have no
 problems fitting small features of the data (e.g. in the rotation
 curve). Moreover, contrary to the two-integral models, they can fit
 simultaneously both data sets (ground-based and HST).

To determine the exact values of M/L and BH mass, we compute a grid of
 models in (M/L, BH) and use $\chi^2$-statistics to assign confidence
 levels (see Figure 2). On the basis of the complete data set, models
 without BH can be ruled out at a confidence level $>$ 99.73 \%. The
 best fit values for a (M/L, BH) are $(6.3^{+0.5}_{-0.4} \; \Msun/\Lsun,
 \; 3.0^{+1.7}_{-1.0} \times 10^8 \Msun)$: model B in Figure
 2. Therefore in this galaxy, the ratio BH mass to bulge mass ($M_{\rm
 BH}/M_{\rm bulge}$) is the highest currently known (2.6\%), together with 
the case of NGC 3115 (see e.g., Ho 1998). 

\begin{figure}
\vskip-2truecm
\plotone{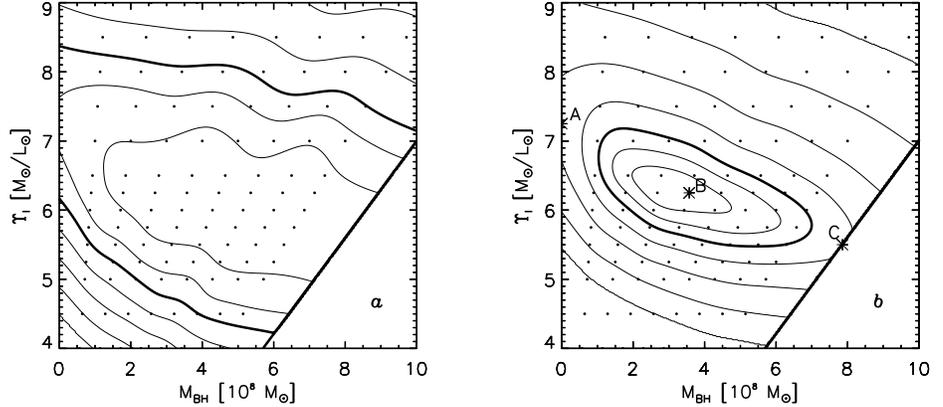}
\vskip-0.5truecm
\caption{Contour plots of iso-$\chi^2$ levels.  The first three
contours define the formal $68.3$, $95.4$ and $99.73$ (thick contour)
per cent confidence levels. Solid dots indicate actual model
calculations.  The $\chi^2$ surface of panel $a$ results from
excluding $h_3$ and $h_4$ as well as all HST/FOS measurements from the
constraints. Panel $b$ shows the $\chi^2$ plot when all
constraints (HST and ground-based) are used.  The
asterisks labeled A, B and C indicate special models discussed in
Section 4.}
\end{figure}

\begin{figure}
\vskip-2cm
\plotone{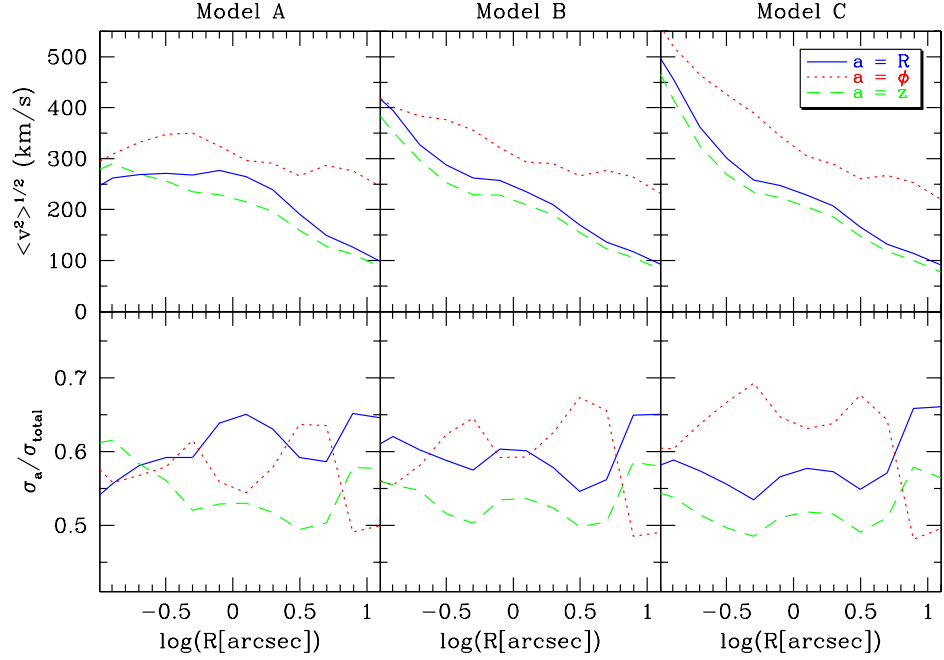}
\vskip-0.5cm
\caption{Internal kinematics in the equatorial plane for models A, B
and C. The first lines shows the second moment and the bottom line the
normalized velocity dispersion $\sigma_a / \sigma_{\rm total}$.  The
insert in the upper right panel explains the line convention.}
\end{figure}

\vskip-3cm
\section{Internal kinematics}

Figure 3 shows the second order velocity moments and velocity
dispersions of models A, B and C averaged over a cone of opening angle
15\deg $\;$around the equatorial plane. This region contains the inner
and outer discs, but in the very center, the bulge contribution
dominates the light.

Models A, B and C mainly differ in the center (due to different BH
masses).  Outside $3''$, $\langle v_{\phi}^2 \rangle$ dominates the
dynamics in accordance with the rapid rotation of the outer disk.
From $3''$ to $12''$, we pass from the {\it azimuthally} anisotropic bulge
into a {\it radially} anisotropic region: the dynamically cold outer
disk built up of close-to-circular orbits with low $\sigma_{\phi}$.

Models B and C have $\sigma_z / \sigma_R$ remarkably constant at $\sim
0.9$, and are thus not too different from two-integral models (for
which this ratio is exactly 1.0).  The same has been observed in M32,
one of the few galaxies for which three-integral models have been
constructed (van der Marel et al. 1998). Merritt and collaborators
(see this volume) have proposed that, under the influence of a central
BH, box orbits are destroyed and the global mass distribution evolves
towards axisymmetry (see also Gerhard \& Binney 1985). At the same
time, this tends to erase any dependence on a third integral in the
DF, consistent with what we have found here. When $M_{\rm
BH}/M_{\rm bulge}$ is $\ge 2.5 \%$, the very short ($\sim$~one
crossing time) evolution towards axisymmetry provides a negative
feedback mechanism that limits the BH mass by cutting off its fuel
suply (Merritt \& Quinlan 1998). Therefore in this scenario, the
maximum BH mass that can be accreted in the center is $2.5 \% \;
M_{\rm bulge}$, close to the observed maximum.

The main change going from model~A to model~C, is a strong increase of
$\sigma_{\phi}/\sigma_R$ in the inner $\sim 3''$. In this region the
circular velocities increase strongly with increasing BH mass.
Nevertheless, all three models provide an almost equally good fit to
the observed rotation velocities because, going from model A to C, the
ratio of $-L_z$-orbits over $+L_z$-orbits increases in the interval
$0.5''~<~R~<~3.0''$. This causes the net streaming motions of all three
models to be roughly similar despite the large differences in circular
velocities and explains the strong increase in
$\sigma_{\phi}$.

\end{document}